\documentclass[draft]{elsart}
\journal{arxiv. LPTENS-06/29.}
\usepackage[latin1]{inputenc}
\usepackage[english]{babel}
\usepackage{amssymb}

\begin{document}

\begin{frontmatter}

\title{Some properties of meta-stable supersymmetry-breaking vacua in 
Wess-Zumino models}

\author{S\'ebastien Ray}

\ead{sebastien.ray@lpt.ens.fr}

\address{Laboratoire de physique th\'eorique de l'\'Ecole normale
	sup\'erieure\\
24 rue Lhomond, 75231 Paris Cedex 05, France}

\begin{abstract}

As a contribution to the current efforts to understand 
super\-symmetry-breaking by meta-stable vacua, we study general properties 
of super\-symmetry-breaking vacua in Wess-Zumino models: we show that 
tree-level degeneracy is generic, explore some constraints on the 
couplings and present a simple model with a long-lived meta-stable vacuum, 
ending with some generalizations to non-renormalizable models.

\end{abstract}


\end{frontmatter}

\section*{\normalsize Introduction}

In the search for a natural model of dynamical supersymmetry breaking, it 
was suggested by Intriligator \emph{et al.} \cite{Intriligator:2006dd} 
that supersymmetry need not be broken by a stable vacuum and that the 
non-supersymmetric vacuum could be a long-lived meta-stable vacuum, with 
possible but slow tunnelling towards a stable, supersymmetric vacuum. This 
idea has recently attracted much attention 
\cite{Banks:2006ma,Argurio:2006ew} since it gives more freedom to 
dynamical supersymmetry breaking, removing for instance the Witten index 
constraint. It could also be of some interest with respect to possible 
embeddings in string theory 
\cite{Braun:2006da,Ooguri:2006pj,Franco:2006es} or M theory 
\cite{Braun:2006em}.

This note presents some simple remarks on the possibility of meta-stable 
super\-symmetry-breaking vacua in \'O~Raifeartaigh-like models. Although 
some of these may be known to experts, they have not, to our knowledge, 
appeared in literature, and they could be useful in coming efforts to 
build realistic models.

We first present some properties of supersymmetry-breaking vacua in 
renormalizable Wess-Zumino models: they are necessarily degenerate at tree 
level and this degeneracy is lifted by a one-loop pseudomodulus 
stabilization. We then study meta-stability due to tunnelling towards a 
neighbouring supersymmetric vacuum and show that the lifetime can be 
parametrically long while leaving the couplings finite. We then try to 
generalize these results to non-renormalizable models.

\section{Renormalizable models}

We consider several chiral superfields $\phi^a$ with canonical K\"ahler 
potential $K=\phi_a^\dagger \phi^a$ and superpotential $W$, third-order 
polynomial of the $\phi^a$.

Suppose there exists a non-supersymmetric vacuum: the potential
$V=|\partial W|^2$  for scalar fields admits a local non-zero minimum. 
The fourth-order expansion of $V$ around the vacuum, exact for a 
third-order superpotential, is then:

\begin{eqnarray}
\label{exp4}
V &=& \left|\partial W\right|^2 + 2\Re\left(\partial^b W^\dagger 
\partial_{ab} W 
\delta\phi^b\right)\nonumber\\
&+& \left|\partial_{ab}W \delta\phi^b\right|^2 + 
\Re\left(\partial^c W^\dagger \partial_{abc} W \delta\phi^a 
\delta\phi^b\right)\\
&+&
\Re\left(\partial^{cd} W^\dagger \partial_{abd} W \delta\phi^a 
\delta\phi^b \delta\phi^\dagger_c\right)
+ \left|\partial_{abc}W \delta\phi^b \delta\phi^c\right|^2,\nonumber
\end{eqnarray}

so that obvious necessary conditions for such a vacuum are:

\begin{equation}
\left\{\begin{array}{l}
\partial W \neq 0\\
\label{eigen}\partial W^\dagger \partial^2 W = 0.
\end{array}\right.
\end{equation}

We shall now try to find some consequences of those conditions in the form 
of constraints on the superpotential.

\bigskip\emph{Degeneracy}

In this paragraph we show that the potential is necessarily exactly 
degenerate at tree level. Using expansion (\ref{exp4}) of the potential at 
vacuum point, we find:

\begin{eqnarray}
\delta V &=& \left|\partial_{ab}W \delta\phi^b\right|^2 + 
\Re\left(\partial^c W^\dagger \partial_{abc} W \delta\phi^a 
\delta\phi^b\right)\nonumber\\
&+& \Re\left(\partial^{cd} W^\dagger \partial_{abd} W 
\delta\phi^a \delta\phi^b \delta\phi^\dagger_c\right) + O(\delta\phi^4).
\end{eqnarray}

This must be positive in order for the vacuum to be (meta)stable. But 
taking $\delta\phi^a = \delta z\partial^a W^\dagger$, with some complex 
$\delta z$, we find, using formula (\ref{eigen}):

\begin{equation}
\delta V = \Re\left((\partial W^\dagger)^3 \partial^3 W \delta 
z^2\right) + O(\delta z^3).
\end{equation}

The first term, if non-zero, is negative for some phase of $\delta z$. As 
the form is supposed to be positive, this yields:

\begin{equation}
\partial_{abc} W \partial^a W^\dagger \partial^b W^\dagger \partial^c 
W^\dagger =0.
\end{equation}

Then if we choose $\delta\phi^a = \varphi^a\delta z^2 + \partial^a
W^\dagger \delta z$, with any $\varphi^a$ such as $\varphi^a \partial_a W
= 0$ and make the same calculation, we find:

\begin{equation}
\delta V = 2\Re\left(\partial^b W^\dagger \partial^c W^\dagger 
\partial_{abc} W \varphi^a \delta z^3\right) + O(\delta z^4).
\end{equation}

At leading order in $\delta z$, positivity implies:

\begin{equation}
\partial^b W^\dagger \partial^c W^\dagger \partial_{abc} W \varphi^a = 0.
\end{equation}

As this is true for any $\varphi^a$ orthogonal to $\partial^a W^\dagger$ 
and for $\partial^a W^\dagger$ itself, this gives:

\begin{equation}
\partial^b W^\dagger \partial^c W^\dagger \partial_{abc} W = 0.
\end{equation}

From this we infer that, for a finite shift in the $\partial W$ direction, 
$\Delta\phi^a = z\partial^a W^\dagger$,

\begin{equation}
\Delta V=0.
\end{equation}

In other words, the potential is degenerate in the $\partial W$ direction.

\bigskip\emph{Coupling conditions}

If we choose the considered supersymmetry-breaking vacuum to be at 
$\phi^a=0$ and the direction $\phi^0\equiv X$ to be the direction of 
$\partial W^\dagger$, the orthogonal directions being labelled by indices 
$i,j,...$, the superpotential, given the previous result, can be written 
as follows:

\begin{equation}
\label{general}
W=\xi X + \frac{1}{2}\left(\mu_{ij}+\lambda_{ij}X + 
\frac{1}{3}\lambda_{ijk}\phi^k\right)\phi^i\phi^j,
\end{equation}

with $\xi$ a real positive number parametrizing the amount of 
supersymmetry breaking. The vacuum extends on the complex line $\phi^i=0$, 
with $X$ taking any value. Instead of keeping the background $\langle 
X\rangle$ as a free parameter, we shall shift it to zero by a change of 
$\mu$.

The masses of the bosonic and fermionic fields around that vacuum are
generically given by the eigenvalues of the following mass matrices:

\begin{equation}
\left\{\begin{array}{l}
M_0^2=\left(\begin{array}{cc}\partial^2W^\dagger \partial^2W & 
\partial^3W^\dagger \partial W \\ \partial W^\dagger \partial^3 W & 
\partial^2 W \partial^2 W^\dagger\end{array}\right),\\
M_{1/2}^2=\left(\begin{array}{cc}\partial^2W^\dagger \partial^2W & 0 \\ 
0 & \partial^2 W \partial^2 W^\dagger\end{array}\right).
\end{array}\right.
\end{equation}

In this case:

\begin{equation}
M_0^2=\left(\begin{array}{cccc}0&0&0&0\\
0&\mu^\dagger\mu & 0&\xi\lambda^\dagger \\
0&0&0&0\\
0&\xi\lambda & 0&\mu \mu^\dagger\end{array}\right) , \;
M_{1/2}^2=\left(\begin{array}{cccc}0&0&0&0\\
0&\mu^\dagger\mu & 0&0 \\
0&0&0&0\\
0&0 & 0&\mu \mu^\dagger\end{array}\right).
\end{equation}

The zero lines and columns correspond, in the bosonic part, to the 
complex, classically massless, $X$ direction and, in the fermionic part, 
to its superpartner the goldstino. They can be left out of the matrices, 
thus giving:

\begin{equation}
M_0^2=\left(\begin{array}{cc}
\mu^\dagger\mu & \xi\lambda^\dagger \\
\xi\lambda & \mu \mu^\dagger\end{array}\right) , \;
M_{1/2}^2=\left(\begin{array}{cc}
\mu^\dagger\mu & 0 \\
0 & \mu \mu^\dagger\end{array}\right).
\end{equation}

Note that the couplings $\lambda_{ijk}$ play no role in the mass terms 
around this line of vacua. The background $\phi=0$ is a vacuum only if the 
matrix $M_0^2$ is positive. This condition can be written:

\begin{equation}
\forall\psi_1,\psi_2 , \, \|\mu\psi_1\|^2 + \|\mu^\dagger\psi_2\|^2 
+ 2\xi\Re\left(\psi_2^\dagger\lambda\psi_1\right) \geq 0.
\end{equation}

Suppose $\det\mu=0$: as a symmetrical matrix, $\mu$ can be written in a 
certain basis of the fields:

\begin{equation}
\mu=\left(\begin{array}{cc}\tilde\mu&0\\0&0\end{array}\right),
\end{equation}

with $\det\tilde\mu\neq0$. Then, with

\begin{equation}
\lambda=\left(\begin{array}{cc}\tilde\lambda&\Lambda\\
^t\Lambda&\tilde\lambda'\end{array}\right), \, 
\psi=\left(\begin{array}{c}\tilde\psi\\\tilde\psi'\end{array}\right), 
\end{equation}

the positivity condition becomes:

\begin{eqnarray}
2\xi\Re\left(\tilde\psi_2^\dagger\tilde\lambda\tilde\psi_1
+\tilde\psi_2'^\dagger{}^t\Lambda\tilde\psi_1
+\tilde\psi_2^\dagger\Lambda\tilde\psi_1'
+\tilde\psi_2'^\dagger\tilde\lambda'\tilde\psi_1'
\right) \nonumber\\
\quad+ \|\tilde\mu\tilde\psi_1\|^2 + \|\tilde\mu^\dagger\tilde\psi_2\|^2 
\geq0.
\end{eqnarray}

This implies $\tilde\lambda'=0=\Lambda$: the primed directions are
massless and can be removed from the calculation. We shall then consider
that $\det\mu\neq0$. The positivity condition is then equivalent to:

\begin{equation}
\forall\psi_1,\psi_2 , \, \|\psi_1\|^2 + \|\psi_2\|^2 
+ 2\xi\Re\left(\psi_2^\dagger U\psi_1\right) \geq 0,
\end{equation}

with $U\equiv\mu^{-1}\lambda\mu^{-1}$. Taking an eigenvalue $u$ of 
$U$ and a corresponding eigenvector $U\psi_1=u\psi_1$ with 
$\|\psi_1\|=1$, then choosing $u^*\psi_2=-|u|\psi_1$, we find:

\begin{equation}
1-\xi|u| \geq 0.
\end{equation}

To put it in words, the positivity condition implies that all 
eigenvalues of $U$ have modules inferior to $\xi^{-1}$:

\begin{equation}
\det\left(\mu^{-1}\lambda\mu^{-1} - u\right)=0 \Rightarrow |u|\leq 
\xi^{-1},
\end{equation}

or equivalently

\begin{equation}
\label{contrap}
0<|v|<\xi \Rightarrow \det\left(\mu^2 - v\lambda\right) \neq0.
\end{equation}

\bigskip\emph{Tree-level stability}

In order for the vacuum to be effectively stable, the matrix $M_0^2$ has 
to be positive on the whole complex line $X$. The coupling $\mu(X)$ is 
equal to $\mu + \lambda X$, so that, using condition (\ref{contrap}) for 
stability, we find:

\begin{equation}
\label{contrapg}
0<|v|<\xi \Rightarrow \det\left[(\mu+X\lambda)^2 - v\lambda\right] 
\neq0\quad \forall X.
\end{equation}

This is a strong condition since the determinant, being a holomorphic 
function of $X$, has no complex root and is thus a constant:

\begin{equation}
\label{trace}
0<|v|<\xi \Rightarrow \partial\det\left[(\mu+X\lambda)^2-v\lambda\right] 
=0
\end{equation}

for all $X$ once again, where $\partial$ stands for the derivative with 
respect to $X$. As this is a holomorphic function of $v$, it must be zero 
for all $v$, so that $\det((\mu+X\lambda)^2-v\lambda)$ is only a function 
of $v$, from which we deduce that the eigenvalues of 
$\mu^{-1}\lambda\mu^{-1}$ are the same all along the complex line $X$.

Expanding equation (\ref{trace}) in powers of $v$, we find, for all 
$n\geq0$ and for all $X$:

\begin{equation}
\label{tr}
\mbox{tr}\{[(\mu+X\lambda)^{-2}\lambda]^n (\mu+X\lambda)^{-1} 
\lambda\}= 0.
\end{equation}

This is not an obviously solvable condition although, for $n=0$, it is
equivalent to say that $\mu^{-1} \lambda$ is nilpotent. We shall solve it
in the following simple case.

\bigskip\emph{Renormalizable three-field model}

If there are only three superfields fields $X,\phi^1,\phi^2$, the matrices 
in question are $2\times2$ and the solutions to the nilpotence condition 
can be written, in a certain basis:

\begin{equation}
\mu=\left(\begin{array}{cc}\mu'&\mu\\\mu&0\end{array}\right),\quad
\lambda=\left(\begin{array}{cc}\lambda&0\\0&0\end{array}\right).
\end{equation}

$\mu'$ can even be set to zero by a shift of $X$, and a phase shift
of $\phi_1$ and $\phi_2$ can be used to make $\mu$ and $\lambda$ real
positive. Condition ($\ref{contrapg}$) then gives:

\begin{equation}
0<|v|<\xi \Rightarrow \mu^2(\mu^2-v\lambda)\neq0,
\end{equation}

i.e. $\mu^2 \geq \xi\lambda$. The general three-field treel-level stable 
renormalizable superpotential with a supersymmetry-breaking vacuum is 
then:

\begin{equation}
\label{oraf}
W=\xi X+\mu\phi^1\phi^2+\frac{1}{2}\lambda X(\phi^1)^2 + 
\frac{1}{6}\lambda_{ijk}\phi^i\phi^j\phi^k,
\end{equation}

with $i,j,k=1,2$. This is none other than the usual \'O~Raifeartaigh model
with an additional $\lambda_{ijk}$ interaction term that is irrelevant for
mass calculation. The masses in the background $\phi^i=0$, apart from the
zero-mass complex particle corresponding to the flat $X$ direction and
from the goldstino, can then be calculated from the mass matrix:

\begin{eqnarray}
\label{mb}
m_B^2 &=& \frac{1}{2}\left[2\mu^2+\lambda^2|X|^2+\epsilon\xi\lambda
\nonumber\right.\\
&& \quad\left.\pm \sqrt{(\lambda^2|X|^2 + \epsilon\xi\lambda)^2 + 
4\mu^2\lambda^2|X|^2}\right],\\
\label{mf}
m_F^2 &=& \frac{1}{2}\left[2\mu^2+\lambda^2|X|^2 \right.\nonumber\\
&& \quad\left.\pm \sqrt{\lambda^4|X|^4 + 4\mu^2\lambda^2|X|^2}\right],
\end{eqnarray}

where $\epsilon^2=1$. These masses are all positive given the condition
$\mu^2 \geq \xi\lambda$.

\bigskip\emph{One-loop stability}

The stabilization of the pseudo-modulus by one-loop potential lifting is a 
well-known result, which can be found for instance in appendix A of 
\cite{Intriligator:2006dd}. We recall here the main lines of the 
calculation, with an additional check that the potential is non-tachyonic 
at infinity. The one-loop correction to the vacuum energy is given in 
model (\ref{oraf}) by:

\begin{equation}
V_{eff}=\frac{1}{64\pi^2}\hbox{Str}\left(M^4 
\ln\frac{M^2}{\Lambda^2}\right).
\end{equation}

That energy, once again, does not depend on the $\lambda_{ijk}$ couplings.  
It is not \emph{a priori} independent from the modulus $X$ and thus
generates a correction potential along that direction. This could make the
vacuum instable if the potential develops tachyonic directions. But if the
correction is positive for $|X|\rightarrow\infty$, then there exists at
least one potential minimum that will be a (meta)stable vacuum.

In the much constrained model considered above, the potential for 
$|X|\rightarrow\infty$ is easily calculated since we have the expressions 
(\ref{mb}) and (\ref{mf}) for the masses. These expressions, for the plus 
sign of $\pm$, give:

\begin{eqnarray}
m^4\ln\frac{m^2}{\Lambda^2} &\simeq& \left[\lambda^4|X|^4 + 
2\lambda^2|X|^2 (2\mu^2 + \epsilon\xi\lambda)
+ 2\mu^4+2\epsilon\mu^2\xi\lambda+\epsilon^2\xi^2\lambda^2
\right]\nonumber\\
&&\times\left[\ln\frac{\lambda^2|X|^2}{\Lambda^2} + \frac{2\mu^2 + 
\epsilon\xi\lambda}{\lambda^2|X|^2}
- \frac{6\mu^4 + 6\epsilon\mu^2\xi\lambda + 
\epsilon^2\xi^2\lambda^2}{2\lambda^4|X|^4} \right],
\end{eqnarray}

where $\epsilon=\pm1$ for bosons and $\epsilon=0$ for fermions. Thus only
$\epsilon^2$ and upper terms contribute (as 2) to the supertrace. The
dominant term for the plus contributions to the supertrace is therefore
$2\xi^2\lambda^2\ln|X|^2$. As for the minus terms, they are of order
$|X|^{-2}$ and therefore negligible, so that:

\begin{equation}
V_{eff}\sim \frac{2\xi^2\lambda^2}{64\pi^2}\ln|X|^2.
\end{equation}

As this is positive for $|X|\rightarrow\infty$ and as the potential is
everywhere well defined, it must admit one or several minima, so that
there exists (meta)stable vacua with no tachyon at one loop\footnote{Note
that such a vacuum might exist without need of tree-level stability on the
\emph{whole} complex plane of the pseudo-modulus---a sufficient condition
would be tree-level stability in a region around the vacuum; however,
after fruitlessly searching for a counter-example, we conjecture that, in
renormalizable models, local one-loop stability is only achieved in models
with global tree-level stability.}.

In fact, $X=0$ is such a minimum: expanding the masses around $X=0$, the 
potential reads:

\begin{equation}
V_{eff}\simeq V_0(\Lambda) + \frac{\lambda^2\mu^2|X|^2}{32\pi^2} F(x)
+ O(|X|^4),
\end{equation}

\begin{equation}
F(x)\equiv \frac{1+x^2}{x}\ln\left(\frac{1+x}{1-x}\right) + 
2\ln(1-x^2) -2,
\end{equation}

with $x\equiv \lambda\xi/\mu^2 \leq1$. As the bracketed function of $x$ 
---call it $F(x)$--- is always positive, with $F(0)=0$ and $F(1)=4\ln2$, 
the model has a (meta)stable vacuum at $X=0$. The bosonic masses at that 
point are $\mu^2$ (complex), $\mu^2 (1\pm x)$ (real) and $\mu^2\lambda^2 
F(x)/32\pi^2$ (complex, one-loop light mass) ; the fermionic masses are 
$\mu^2$ (two Weyl spinors) and of course exactly zero for the Goldstino.

\bigskip\emph{Meta-stability and supersymmetric vacua}

After studying the local behaviour of the system around the 
supersymmetry-breaking vacuum, we shall now give some hints of possible 
non-perturbative effects due to other vacua. In the general model 
(\ref{general}) as well as in the three-field model (\ref{oraf}), the 
vacuum is not generically unique, and in particular, there exists 
generically a supersymmetric vacuum, except, for instance, if we impose 
some global symmetry on the model \cite{Nelson:1993nf}. The vacuum will 
then be meta-stable, tunnelling towards a more stable, generally 
supersymmetric, vacuum.

For simplicity, let us study the \'O~Raifeartaigh-like model. Besides the
studied vacuum, there will generically be four supersymmetric vacua; in
special cases there can be less of them (three or two) or even a whole
complex line of degenerate supersymmetric vacua. In other special cases,
as the original \'O~Raifeartaigh model, there will not be any possible
supersymmetric vacuum, but several supersymmetry-breaking vacua.

Let us present a simple model of meta-stable supersymmetry-breaking vacuum 
tunnelling to a supersymmetric vacuum. We shall change variables for 
simplicity and write it:

\begin{equation}
W=h\left[\Phi^2\phi_1 - m\Phi(\phi_1+\alpha\phi_2)\right],
\end{equation}

with $\alpha^2 < 1/8$. This model admits a degenerate supersymmetric 
vacuum at $\Phi=0$, $\phi_1+\alpha\phi_2$, and a meta-stable 
supersymmetry-beaking vacuum at $\Phi=(3+\sqrt{1-8\alpha^2})/4\times m$, 
$(m-2\Phi)\phi_1 + \alpha m\phi_2 = 0$. The latter is stabilized at one 
loop for $\phi_1=\phi_2=0$.

The lifetime of the meta-stable vacuum can be easily evaluated: as the
model has a $U(1)_R$ symmetry under which $\Phi$ is neuter and $\phi_1$
and $\phi_2$ have charge 2, the plane $\phi_1=\phi_2=0$ is stable under
the equations of motion and the least potential barrier path will involve
only $\Phi$ changes. Moreover, for $\phi_1=\phi_2=0$, the potential is
invariant under $\Phi\rightarrow\Phi^*$, so that, as the meta-stable value
of the field is real, it remains so during the bounce. One can then write
the potential as a function of a real $\Phi$:

\begin{equation}
V(\Phi)=h^2\left[\Phi^4 - 2m\Phi^3 + m^2(1+\alpha^2)\Phi^2\right].
\end{equation}

We can then, for small $\alpha$, use the known results on the behaviour of 
false vacua \cite{Callan:1977pt} in the thin-wall approximation, for which 
the energy density difference between the two vacua, here $h^2 m^4 
\alpha^2 + O(\alpha^4)$, is small. At leading order in $\alpha^2$, the 
probability of tunnelling per unit time per unit volume then reads:

\begin{equation}
\Gamma/V \propto\exp\left[-\frac{\pi^2}{24h^2\alpha^6}\right],
\end{equation}

with an average radius of the tunnelling region:

\begin{equation}
\bar\rho = \frac{1}{\sqrt{2} h m \alpha^2}.
\end{equation}

The lifetime of the meta-stable, supersymmetry-breaking vacuum is thus 
parametrically great in the limit where field $\phi_2$ decouples and the 
supersymmetry-breaking scale is small.

\section{\normalsize Non-renormalizable generalizations}

We shall now try to extend the previous results to non-renormalizable 
models, i.e. higher-order superpotentials and non-canonical K\"ahler 
potentials.

\bigskip\emph{Degeneracy for canonical K\"ahler}

The degeneracy theorem for renormalizable models is easily extended to 
general superpotentials, provided we keep a canonical K\"ahler potential: 
it can be shown that a non-zero minimum of a potential of the form 
$V=|\partial W|^2$ is always perturbatively degenerate.

We shall then use a recurrence to show that the potential at 
supersymmetry-breaking vacuum point is flat at all orders in the $\partial 
W^\dagger$ direction. Let us use the convention $A_k \equiv (\partial 
W^\dagger)^k \partial^{k+1} W$, where $k$ of the indices of the multiple 
derivative are contracted with the $k$ simple derivatives. The vacuum 
conditions can then be written $A_0 \neq 0$, $A_1 = 0$.

Let us now suppose, as a recurrence condition, that for some non-zero
integer $n$, $A_k=0\; \forall1 \leq k \leq n$. Let us then consider a
variation of the fields $\phi^i$ around the vacuum $\delta\phi^i =
\partial^i W^\dagger\delta z + \varphi^i\delta z^{n+1}$, with $\varphi^i
\partial_i W=0$. The leading term of the variation of $V$ for small
$\delta z$ must be positive whatever the choice of the direction
$\varphi^i$.

For $1\leq k\leq n$, the $k$-th order of variation of $V$ in $\delta 
z$ reads:

\begin{equation}
\delta^k V = \sum_{i=0}^k \frac{\delta z^i \delta\bar z^{k-i}}{i! 
(k-i)!} A_{k-i}^\dagger A_i = 0
\end{equation}

by recurrence condition. Furthermore, for $0\leq k\leq n$, the
$(n+k+1)$-th order reads:

\begin{eqnarray}
\delta^{n+k+1} V &=& \sum_{i=0}^{n+k+1} \frac{\delta z^i \delta\bar 
z^{n+k-i+1}}{i! (n+k-i+1)!} A_{n+k-i+1}^\dagger A_i\\ \nonumber
&&+ 2\Re\left\{ \sum_{i=0}^k
\frac{\delta z^{n+i+1} \delta \bar z^{k-i}}{i! (k-i)!} A_{k-i}^\dagger 
\left[\varphi (\partial W^\dagger)^i \partial^{i+2} W \right]\right\}\\
 &=& 2\Re\left\{\delta z^{n+k+1}\left[\frac{1}{(n+k+1)!} A_0^\dagger 
A_{n+k+1} + \frac{1}{k!} \varphi A_{k+1}\right]\right\}.
\end{eqnarray}

These terms must all be zero since, if one of them were not, the leading
order in $\delta z$ would be of the form $\Re(\delta z^{n+k+1})$, which
takes negative values for some $\delta z$. Hence:

\begin{equation}
\frac{1}{(n+k+1)!} A_0^\dagger A_{n+k+1} = -\frac{1}{k!} \varphi A_{k+1}.
\end{equation}

For $k=0$, this gives us $A_0^\dagger A_{n+1}=0$ and, since the
equation must hold for every initial choice of the direction $\varphi$,
it yields, for $k=n$, $\varphi A_{n+1}=0$. From these two results we
finally conclude that $A_{n+1}=0$: the recurrence condition is verified
one step further. An additional result, if we take $\varphi=0$, is that
the potential in the $\partial W^\dagger$ direction is flat up to order
$2n+1$. As the recurrence condition is true for $n=1$, it is true for
all $n$, and the potential is flat at all orders.

Thus for a canonical K\"ahler potential and an analytic superpotential,
a super\-symmetry-breaking vacuum is always degenerate since for any 
complex $z$, $V(\phi_0 + z\partial W^\dagger)=V(\phi_0)$.

\bigskip\emph{Non-canonical K\"ahler potentials}

That theorem only holds for a canonical K\"ahler potential: for a generic 
K\"ahler potential, the vacuum need not be degenerate at all, as is 
obvious from the following one-superfield counter-example:

\begin{eqnarray}
K&=& \phi^\dagger\phi - \frac{1}{4m^2}(\phi^\dagger\phi)^2,\\
W&=& h\left[ \frac{\alpha^2(3-\alpha^2)}{2}m\phi^2 - \alpha\phi^3 + 
\frac{1}{4m}\phi^4 \right],
\end{eqnarray}

where $\alpha < 1$. The lagrangian density for the scalar part of this 
theory is:

\begin{eqnarray}
L&=&\left(1-\frac{1}{m^2}|\phi|^2\right) \partial_\mu\phi^\dagger 
\partial^\mu\phi \nonumber\\&-& \frac{h^2|\phi|^2}{m^2-|\phi|^2} 
\left|\phi^2 - 3\alpha m\phi + \alpha^2(3-\alpha^2) m^2\right|^2.
\end{eqnarray}

There is a supersymmetric vacuum at $\phi=0$ and a meta-stable, 
non-denegerate supersymmetry-breaking vacuum at $\phi=\alpha m$, the mass 
of the scalar (complex) particle around that vacuum being $h\alpha^3 
m/(1-\alpha^2)$: there is no pseudo-modulus even at tree level. As 
$\alpha\rightarrow 0$, the meta-stable vacuum becomes long-lived, 
according to the thin-wall approximation model.

Non-renormalizable models are therefore far less constrained as regards
the properties of their vacua and it seems difficult to characterize them
by generic features. Long-lived meta-stable vacua are still easily found,
as the previous example shows, but still require some fine tuning in the
couplings.

\section*{\normalsize Conclusion}

This paper aimed to be a modest exploration of the properties of 
meta-stable supersymmetry breaking in non-gauged Wess-Zumino-like 
theories---degeneracy and modulus stabilization. In order for this type of 
$F$-term breaking to be transplanted in a realistic theory, the essential 
problem would be the one-loop light mass of the pseudo-modulus, yielding 
an unobserved light scalar in addition to the generic massless fermion.

\bigskip\emph{Acknowledgements}

I am much indebted, for discussions, advice and encouragements, to Costas 
P. Bachas.

\end{document}